\documentstyle[titlepage,12pt]{article}
\begin{document}
\parindent 2em

\begin{titlepage}
\begin{center}
\vspace{12mm}
{\LARGE Reply to comment on "Continuum dual theory of the transition 
in 3D lattice superconductor"} 
\vspace{15mm}

Igor F. Herbut\\

Department of Physics and Astronomy, University of British Columbia, 
6224 Agricultural Road, Vancouver B. C., Canada V6T 1Z1\\

\end{center}
\vspace{10mm}
{\bf Abstract:}
It is argued that the penetration depth and the correlation 
length at the critical point of the 3D superconductor 
diverge with the same critical exponents, as follows from the general 
scaling arguments and from the independent calculations in 
both Ginzburg-Landau and its dual theory. The incorrect 
prediction of Kiometzis, Kleinert and Schakel (KKS) that this is 
not so is the result of a faulty treatment of the 
continuum version of the dual theory, in which two, 
instead of one, coupling constants are tuned to reach 
the critical point. The recent paper by the present author 
criticized by KKS in the preprint cond-mat/9702159 differs 
on this point from KKS, and consequently obtains 
the expected relation
between the penetration depth and the correlation length
in the critical region.  
\end{titlepage}


In a recent paper \cite{herbut1}, the present 
author has rederived the continuum dual 
theory for the  3D Ginzburg-Landau superconductor 
(scalar electrodynamics) starting from 
the lattice electrodynamics, and studied its critical behavior.  
It was shown that 
the dual vector-field does not influence the critical behavior of the 
disorder field, which is therefore in the universality class 
of the 3D XY model. This implies that the correlation length 
for the disorder field, which on general grounds is 
expected to be the only diverging 
length-scale in the problem, diverges at the transition with 
the XY exponent, as well known \cite{dasgupta}. It was further 
demonstrated that the stiffness $K$ for the phase fluctuations of the 
original order-parameter (describing Cooper pairs) vanishes at the 
transition with the  power in agreement with the Josephson relation 
in 3D, as follows from the identification of the 
correlation length for the disorder field with the single 
diverging length scale in the problem. This result, when 
combined with the exact \cite{herbut2} result for the anomalous 
dimension of the original gauge-field implies that 
the penetration depth for the physical magnetic field diverges with 
the same exponent as the correlation length. The comparison 
with the recent perturbative \cite{herbut2} and non-perturbative 
\cite{bergerhof} calculations of the critical exponents 
in the original 3D scalar electrodynamics  was discussed. 

   Studying the original 3D scalar electrodynamics, the author 
with Tesanovic has recently demonstrated \cite{herbut2}
that at the infrared-stable fixed point 
of the theory, the penetration depth and the correlation length 
for the order-parameter in the superconducting gauge diverge 
with the same exponent. Their ratio, defined as the 
renormalized Ginzburg-Landau parameter $\kappa$, in the critical region 
approaches a universal constant, which is a characteristic 
of the fixed point, similarly to the critical exponents. 
This is precisely what should be expected, since at the 
fixed point in question there 
is only one relevant direction in the RG sense, 
the direction of temperature, 
and therefore a single diverging length-scale. A problem 
treated in ref. 1 is how to address the same question in the dual 
formulation of the problem, which is completely expressed 
in terms of auxiliary fields (disorder field and the dual 
vector-field), while the original physical fields are integrated 
out. For that purpose it proved useful to trace the 
appearance of the stiffness $K$ in the original and the dual 
theories. Since the power-law vanishing of $K$ can be easily 
deduced from the dual theory, and $K$ is simply related to the 
physical penetration depth, 
the requisite divergence of the penetration depth 
readily followed, and not surprisingly, 
 it agreed with the expectation that it must be proportional 
to the diverging correlation length. The result 
also seems to agree with an early study of the lattice 
superconductor by Peskin \cite{peskin}. 
The crucial step in arriving at this 
conclusion was the inclusion of the anomalous dimension for the 
original massless gauge-field. The dual theory studied in 
ref. 1 alone can not 
account for this, and this information has to be supplied externally.
This is not at all surprising, since the correlation functions 
of the auxiliary 
fields appearing in the dual theory are in no simple relation
to the correlation functions of the original physical fields.
  
The result of Kiometzis, Kleinert and Schakel (KKS) \cite{kiometzis} 
which claims that the penetration depth diverges with the mean-field 
exponent is wrong on 
both conceptual and technical level. First, it implies two 
differently diverging length-scales at the transition, in violation of 
the above described results in the original scalar 
electrodynamics and of the general scaling arguments. 
Second, as discussed in ref. 1, {\it assuming} a mean-field 
form (or any other form) 
for vanishing of the mass of the dual vector field at some 
preset transition temperature is obviously incorrect. 
An immediate question arises: at which transition temperature?
 Since there are no additive 
renormalization of the vector-field 
 mass in the theory, it will really vanish at 
the assumed transition temperature, and not at the 
shifted, true transition temperature, which, of course, can not be known 
in advance to be put in instead. The point is that even 
in the dual theory the transition is tuned by varying a single parameter, 
namely, the mass of the disorder field, while all other coupling 
constants in the theory, like the mass of the dual vector-field, 
renormalize accordingly. This is in stark contrast to the procedure 
used by KKS, where in effect two parameters, masses of both disorder 
field and dual vector-field, must be tuned by hand to arrive at the 
critical point.

  The apparent experimental support \cite{lin} for the KKS result must be 
taken with reservation for several reasons. First, their result is 
in fact in better agreement with the 
non-trivial exponent derived by the 
author and Tesanovic \cite{herbut2}.
It seems likely, however, that the experiment of Lin et al. is 
not yet in the critical region of the charged fixed point in question.
From the estimates of the Ginzburg number, one would expect that for the 
experimentally studied range of temperatures the system is in the crossover 
region close to the unstable, neutral fixed point (see the flow
diagram in ref. 3 and \cite{kammal}). 

  In conclusion, I argued that the claim that the penetration depth 
and the correlation length diverge exactly the same way at the 
stable fixed point in the theory should be correct on general 
grounds, and that indeed it is found to be correct in both original 
and the dual formulations of the 3D scalar electrodynamics. 
The result of KKS claiming otherwise follows form the incorrect 
treatment of the dual theory.

\end{document}